# Electric-field control of anomalous and topological Hall effects in oxide bilayer thin films

Yuki Ohuchi[1], Jobu Matsuno[2*], Naoki Ogawa[2], Yusuke Kozuka[1], Masaki Uchida[1], Yoshinori Tokura[1,2] and Masashi Kawasaki[1,2]

[1]*Department of Applied Physics and Quantum-Phase Electronics Center (QPEC), University of Tokyo, Tokyo 113-8656, Japan*

[2] *RIKEN Center for Emergent Matter Science (CEMS), Wako, Saitama 351-0198, Japan*

[*]Author to whom correspondence should be addressed. Electronic mail: matsuno@riken.jp

**One of the key technologies in spintronics is to tame spin-orbit coupling (SOC) that links spin and motion of electrons, giving rise to intriguing magneto-transport properties in itinerant magnets. Prominent examples of such SOC-based phenomena are anomalous[1] and topological Hall effects[1–3]. However, controlling them by electric field has remained unachieved since electric field tends to be screened in itinerant magnets. Here we demonstrate that both anomalous and topological Hall effects can be modulated by electric field in oxide heterostructures consisting of ferromagnetic $SrRuO_3$ (ref. 4) and nonmagnetic $SrIrO_3$. We observed clear electric-field effect only when $SrIrO_3$ is inserted between $SrRuO_3$ and a gate dielectric. Our results establish that strong SOC of nonmagnetic materials such as $SrIrO_3$ (ref. 5) is essential in electrical tuning of these Hall effects and possibly other SOC-related phenomena[6–8].**

Profound implications on the electric-field control of spin states have been revealed through magneto-transport properties arising from $SOC^9$. Such properties have been attracting great interest as vital foundations of high-density and low-power-consumption spintronic devices because the properties cooperate on spin manipulation without magnetic-field variation or large current injection. Among the magneto-transport processes, promising is intrinsic anomalous Hall effect (AHE), which is related to magnetization (*M*) in Hall resistivity ($\rho_{AHE} = R_S M$). In some itinerant ferromagnets such as SrRuO$_3$ (refs 4,10), the proportionality factor ($R_S$) is governed by *k*-space monopoles, i.e., singularities originating from band-crossings gapped by SOC; not *M* but $R_S$ is potentially tuneable by electric field there. Another intriguing example is topological Hall effect (THE) generated by magnetic skyrmion[2,3], a topologically protected spin swirling texture[11–13]. Skyrmion has various gifted characteristics as an information carrier in spintronic devices[14]; particularly useful is the nanometre-sized one induced by Dzyaloshinskii-Moriya interaction (DMI), which arises from the combination of SOC and broken inversion symmetry[13,15]. Despite mounting interests in electrical control of these Hall effects, however, their control in itinerant magnets has been elusive in the conventional field-effect structures, that is, magnetic materials adjacent to gate dielectrics. Recent discovery of THE in the oxide heterostructure composed of SrRuO$_3$ and SrIrO$_3$ (ref. 16) brings an opportunity to approach the electrical control because the strong SOC in SrIrO$_3$ (ref. 5) induces a novel SOC-related phenomenon in neighbouring ferromagnetic SrRuO$_3$. In addition to the intrinsic AHE inherent to SrRuO$_3$ (ref. 4), the interface asymmetry gives rise to the substantial DMI and the concomitant THE through skyrmion formation. In view of the fact that ultrathin films of itinerant magnets have been a playground for electrical manipulation of spin states[6,9,17–20], electric field applied to the ultrathin heterostructure may provoke effective manipulation of the spin state contending with screening effects. Furthermore, we can expect that the strong SOC in the vicinity of the interface plays a positive role as well. Here, we show the clear modulations of both AHE and THE by electric field in the structure of SrRuO$_3$/SrIrO$_3$/SrTiO$_3$ (SRO$m$/SIO2/Sub, $m$ = 3 - 5 unit cells, Fig. 1a) from magneto-transport (Fig. 1b) and magneto-optic Kerr effect (MOKE) measurements.

    5-unit-cell SrRuO$_3$ thin films were epitaxially grown on SrTiO$_3$(001) substrates by pulsed laser deposition in three stacking orders with SrIrO$_3$ (see Method): a single layer of SrRuO$_3$ (SRO5/Sub), a bilayer of SrRuO$_3$ and 2-unit-cell SrIrO$_3$ (SIO2/SRO5/Sub), and that with the inverted deposition order (SRO5/SIO2/Sub). In Fig. 1c, we show temperature dependence of longitudinal resistivity ($\rho_{xx}$) and magnetization perpendicular to the film plane in the three samples. All of them show metallic conduction and ferromagnetic magnetization whose Curie temperatures ($T_C$) are lower than the bulk value (160 K)[4]. These transport and magnetic properties are consistent with previously reported ultrathin SrRuO$_3$ films[21]. Figure 1d shows anomalous Hall conductivity ($\sigma_{AHE}$) as a function of magnetization with temperature as a control parameter. The $\sigma_{AHE}$ of the three samples behave the same way; when the magnetization gets larger, the sign of $\sigma_{AHE}$ is inverted from positive to negative. This

sign inversion is consistent with the previous experiments[10,16], manifesting the above-mentioned band anti-crossing singularities of SrRuO$_3$. Conversely, such temperature dependence in AHE implies the potential for electrical control of AHE through the modification of the SOC-induced band.

In contrast to the similarities of the basic transport and magnetic properties among the three samples, clear difference emerges in electric-field effect (see Method). The sums of anomalous and topological Hall resistivities ($\rho_{AHE} + \rho_{THE}$) under applied electric fields are shown in Figs. 1e-1g as a function of external magnetic field ($B$). The $B$ dependence of $\rho_{AHE}$ corresponds to that of magnetization normal to the film plane, whereas $\rho_{THE}$ shows up with the formation of skyrmion as exemplified by the peak at around 3 T in Fig. 1g. The sum of these Hall components is obtained by subtracting the $B$-linear ordinary Hall term ($R_H B$) from the Hall resistivity ($\rho_{yx}$). Both in SRO5/Sub (Fig. 1e) and in SIO2/SRO5/Sub (Fig. 1f), $\rho_{AHE} + \rho_{THE}$ is almost unchanged by the applied electric field. However, $\rho_{AHE} + \rho_{THE}$ is obviously changed in SRO5/SIO2/Sub (Fig. 1g). This striking stacking-order dependence suggests that the electric field to the SrRuO$_3$-SrIrO$_3$ interface causes the noticeable modulation; the SrIrO$_3$ layer is designed to be thinner than the SrRuO$_3$ layer and hence the interface in SRO5/SIO2/Sub is closer to the gate dielectric than that in SIO2/SRO5/Sub. Hereafter, we focus on the control in the SRO5/SIO2/Sub sample.

In the structure of SRO5/SIO2/Sub, the large electric-field modulation gives rise to the sign inversion of AHE without changing temperature. Magnetic-field dependences of $\rho_{AHE} + \rho_{THE}$ under different electric fields at 30 K are shown in each panel of Fig. 2a. Except for the topological Hall term which shows a peak at around 0.8 T, the data are dominated by the anomalous Hall contribution. In particular, we can consider the value at high magnetic field such as 2 T to be totally derived from $\rho_{AHE}$ since all the spins are ferromagnetically aligned without forming any skyrmions. The sign of $\rho_{AHE}$ above the saturation field is inverted from negative to positive when the gate voltage is varied from negative (−180 V) to positive (200 V), while $\rho_{AHE}$ is close to vanish under zero bias. This sign inversion indicates the sign reversal of the proportionality factor $R_S$ because MOKE measurements reveal that only a minor fraction of *M*, less than 10% of the magnetization, is electrically modulated (see Supplementary Information). Such electrical sign inversion has never been observed in plain films of itinerant magnets including SrRuO$_3$, although even larger electric field has been applied to it with an electric double layer transistor[20].

As shown in Fig. 2b, the tendency of AHE change is nearly temperature independent; the positive (negative) bias voltage increases (decreases) $\rho_{AHE}$ regardless of the sign of AHE. Since singularities in the band structure of SrRuO$_3$ originate from band-anti-crossings gapped by SOC, we speculate that the electric-field control of AHE is ascribed to the redistribution of the singularities caused by the variation of SOC. The pronounced controllability in SRO5/SIO2/Sub compared with its absence in SIO2/SRO5/Sub suggests that electric field from SrIrO$_3$ side induces certain modification of SOC in the bilayer even if it has itinerant carrier density as high as $10^{22}$ cm$^{-3}$. This is in sharp contrast to the case of chemically doped EuTiO$_3$ (ref. 22), where carrier-density variation inverts the sign of AHE accompanied by the shift of Fermi energy ($E_F$). The insertion of the SrIrO$_3$ layer makes a more important contribution to the AHE modulation than $E_F$ shift in SrRuO$_3$ does. The controllability may partly rely on the lower carrier density of inserted semimetallic SrIrO$_3$, the order of $10^{19}$ cm$^{-3}$ (ref. 23).

THE is also modulated by electric field in the structure of SRO5/SIO2/Sub. In order to evaluate $\rho_{THE}$ from magneto-transport measurement, $\rho_{AHE}$ has to be subtracted from $\rho_{AHE} + \rho_{THE}$. Since MOKE is proportional to magnetization as anticipated from its perturbative nature and also as experimentally verified in ref. 16, MOKE under applied electric field can be utilized as the reference of $\rho_{AHE}$, which is also proportional to magnetization at a constant temperature. The magnetic-field dependences of Kerr rotation are shown by the broken lines in Fig. 2a. Each Kerr rotation is normalized by $\rho_{AHE}$ at high magnetic field under each applied electric field; the normalized curves represent the $B$ dependence of $\rho_{AHE}$. By subtracting them from $\rho_{AHE} + \rho_{THE}$, we deduced $\rho_{THE}$ (yellow coloured regions in Fig. 2a). Figure 3a shows the electric-field-controlled peak of $\rho_{THE}$ around $B = 0.8$ T. The electric field also tunes the range of magnetic field where finite $\rho_{THE}$ appears; when the positive (negative) gate voltage is applied, $\rho_{THE}$ gets smaller (larger) and the magnetic-field region gets shrunk (enlarged). The variation in peak of $\rho_{THE}$ [$\Delta\rho_{THE} = \rho_{THE} (V_G = 200$ V$) - \rho_{THE} (V_G = -180$ V$)$] normalized by zero-bias value [$\Delta\rho_{THE}/\rho_{THE} (0$ V$)$] is as large as 55%.

We discuss the relation of this THE modulation with skyrmion length scale. In this heterostructure, $\rho_{THE}$ is an outgrowth of ordinary Hall effect under fictitious magnetic field ($b_{eff}$) caused by the non-coplanar spin texture of skyrmion[13]. Since one skyrmion generates one flux quantum ($\Phi_0 = h/e$), $\rho_{THE}$ is described as a function of skyrmion density ($N_{sk}$):

$$\rho_{THE} = PR_H b_{eff} = PR_H N_{sk} \Phi_0 \qquad (1)$$

where $P$ is spin polarization of conduction electrons in SrRuO$_3$. Ordinary Hall coefficient $R_H$ is inversely proportional to carrier density. The change of $P$ or $R_H$ is negligible because $E_F$ in SrRuO$_3$ remains almost unchanged judging from the variation of $\rho_{xx}$ as small as the order of 1% as discussed later. Therefore, the modulation of $\rho_{THE}$ indicates the change of $N_{sk}$. The skyrmion size, estimated by $r_{sk} \cong 1/\sqrt{N_{sk}}$, is varied from 15 nm ($V_G = 200$ V) to 12 nm ($V_G = -180$ V) adopting $P = -9.5\%$ as a typical value of SrRuO$_3$ (ref. 24) and $R_H$ evaluated from the $B$-linear component of Hall resistivity under $V_G = 0$ V at 30 K. Since $r_{sk}$ is proportional to the ratio of ferromagnetic interaction $J$ to DMI $D$, or $J/D$ (ref. 13), the positive (negative) gate voltage should increase (decrease) $J/D$. The controlling knob of $\rho_{THE}$ is thought to be $D$ rather than $J$; the variation of $J$ is negligible because the applied electric field hardly varies $T_C$, which can be evaluated by a kink structure in $\rho_{xx}$ (ref. 4, see Supplementary Information).

Considering that the variation of DMI is attributed to the modification of SOC at the interface, both modulations of AHE and THE are brought about by the same origin; this appears in the temperature dependence. The tendency of $\rho_{THE}$ change is independent of temperature as shown in Fig. 3b. The expansion of the observable magnetic-field region with application of negative electric field is discerned at every temperature below 40 K (Fig. 3c). These indicate that the obtained modulation of THE is independent of the sign of AHE, which is inverted at 30 K. The temperature independent trend of the electric-field effect on THE is in accord with that on AHE (Fig. 2b), implying the common origin of SOC variation. In fact, MOKE measurements indicate that the coercive force ($H_C$) is shifted from 0.73 T ($V_G$ = 200 V) to 0.76 T ($V_G$ = −180 V) at 30 K (see Supplementary Information). This shift of $H_C$ also supports the modification of SOC because the magnetic anisotropy originates from SOC.

The observed modulations of AHE and THE are not simply explained by the variation of carrier density. In order to show this clearly, the variations of AHE [$\Delta\rho_{AHE} = \rho_{AHE}$ ($V_G$ = 200 V) − $\rho_{AHE}$ ($V_G$ = −180 V)] and THE [$\Delta\rho_{THE}/\rho_{THE}$ (0 V)] with different thickness of SrRuO$_3$ films are shown in Figs. 4a and 4b, respectively, as a function of $\rho_{xx}$ variation normalized by zero-bias value [$\Delta\rho_{xx}/\rho_{xx}$ (0 V)] at various temperatures (see Supplementary Information for detailed transport and magnetic properties). $\Delta\rho_{xx}/\rho_{xx}$ (0 V) ranges from 1.2% in SRO5/SIO2/Sub to 6.6% in SRO3/SIO2/Sub at 10 K. Since $\rho_{xx}$ is inversely proportional to carrier density ($n$), the order of magnitude is consistent with the expected electron accumulation by a SrTiO$_3$-back-gate transistor (see Supplementary Information). $\Delta\rho_{AHE}$ in bilayers indicates that the gate-bias control of AHE has the specific tendency regardless of the film thickness. The same plots of SrRuO$_3$ single layers (SRO$m$/Sub) are also depicted for comparison. They have negligible change in AHE, while their $\rho_{xx}$ variation is in the same range with that in bilayers with the same SrRuO$_3$ thickness. As already pointed out, these results demonstrate the importance of inserting the SrIrO$_3$ layer rather than $E_F$ shift in SrRuO$_3$. According to equation (1), $\rho_{THE}$ is proportional not only to $N_{sk}$ but also to $R_H$. We show the calculated variation of $\rho_{THE}$ contributed from $R_H$ i.e. $n$ variation [$\Delta R_H$ i.e. $\Delta n = n(V_G = 200$ V$) - n(V_G = -180$ V$)$] with the red dash-dotted line in Fig. 4b, assuming small $\Delta n$ ($\Delta R_H \propto \Delta n/n^2$ i.e. $\Delta\rho_{THE} \propto \Delta n$) and $V_G$-independent electron mobility. All the modulations in heterostructures are located far above the red dash-dotted line. Therefore, most part of $\rho_{THE}$ is attributed to the change of skyrmion size, i.e. DMI. These pronounced modulations of AHE and THE are both achieved not by carrier-density variation in SrRuO$_3$ but by modification of SOC at the interface of SrRuO$_3$ and SrIrO$_3$.

Recent theoretical calculations[25,26] have revealed that DMI is enormously affected by the singularities in band structures as well as AHE is, although the precise expression is different between them. This has been confirmed in itinerant chiral magnets[27] and a semiconducting Dirac electron system[28]. We speculate that our bilayers have some band deformation by applying electric field to the interface. In both plots in Fig. 4, thinner SrRuO$_3$ films exhibit smaller variation in spite of larger $\Delta\rho_{xx}/\rho_{xx}$ (0 V). This thickness dependence is probably attributed to band structure deformation as observed in the thin limit of SrRuO$_3$ (ref. 29); the density of states near $E_F$ shrinks as SrRuO$_3$ gets thinner.

Since the itinerant-electron systems including $SrRuO_3$ and $SrIrO_3$ have complicated distribution of band-anti-crossings in their band structures[23,30,31], the quantitative evaluation of the electric-field-induced modulation in the heterostructures requires elaborate theoretical investigation. Nevertheless, the present observations clearly point out that the electric field applied to the thin $SrIrO_3$ plays the crucial role in SOC at the interface and even brings about the significant modification of magnetic properties in the neighbouring itinerant ferromagnet. The method of inserting a thin non-magnetic material with strong SOC between a ferromagnet and a gate dielectric is applicable to tuning of many intriguing spin-orbit coupled phenomena such as magnetic anisotropy[6], domain wall motion[7] and DMI iteself[8].

Method

The epitaxial bilayers composed of $SrIrO_3$ and $SrRuO_3$, and single layer films of $SrRuO_3$ were deposited on $SrTiO_3$(001) substrates by pulsed laser deposition using a KrF excimer laser ($\lambda = 248$ nm). The substrate temperatures during the growth of $SrRuO_3$ and $SrIrO_3$ were 730°C and 600°C, respectively, where oxygen partial pressure was 120 mTorr. The laser fluence was 1.2 J/cm$^2$ for $SrRuO_3$ and 2.6 J/cm$^2$ for $SrIrO_3$.

The magnetization data were recorded by a SQUID magnetometer with a magnetic field applied perpendicularly to the film plane and along the magnetic easy axis grown on SrTiO$_3$(001). Magneto-optic Kerr effect was measured with a laser at 690 nm wavelength in polar geometry by using a photoelastic modulator.

Transport properties were measured in Hall bars cut by a diamond wheel saw (1 mm × 2.5 mm) and ultrasonically boned with Al wires. The applied current was 10 μA, which corresponds to $3.6 \times 10^6$ A/m$^2$ for SRO5/SIO2/Sub. Back-gate transistors were fabricated using 0.5-mm-thick SrTiO$_3$ substrates as a gate dielectric and silver paste as a gate electrode at the opposite side of the deposited films. Antisymmetrizations were performed for both the Hall resistivity and the Kerr rotation angle. Ordinary Hall term was subtracted from the Hall resistivity by linear fitting in a higher magnetic-field region.

Acknowledgements

We thank N. Nagaosa, R. Arita, T. Koretsune, and Y. Kaneko for fruitful discussions. This work was partly supported by a JSPS Grant-in-Aid for Scientific Research(S) No. 24226002 Japan, by Japan Society for the Promotion of Science through Program for Leading Graduate Schools (MERIT) (Y.O.), by JSPS Fellowship No. 28·06170 (Y.O.).

Author contributions

Y.O., J.M., and M.K. designed the experiments. Y.O. fabricated films, measured transport properties, and analysed the data. N.O. measured magneto-optic Kerr effect. M.U. and Y.K. contributed to the film fabrication and the transport measurements. Y.O. and J.M. wrote the manuscript. Y.T. and M.K. coordinated the projects. All authors discussed the results.

Competing financial interests

The authors declare no competing financial interests.

**Figure Captions:**

**Figure 1 | Structure and basic physical properties of samples**

**a**, Schematics of the SrRuO$_3$-SrIrO$_3$ bilayer film, where SrIrO$_3$ is inserted between SrRuO$_3$ and SrTiO$_3$ substrate (SRO$m$/SIO2/Sub, $m$ = 3 - 5 unit cells). **b**, Schematics of magneto-transport properties observed in SRO$m$/SIO2/Sub under the application of gate voltage ($V_G$). Anomalous Hall effect (AHE) is generated by magnetization ($M$) while topological Hall effect (THE) is driven by fictitious magnetic field ($b_{eff}$) concomitant with skyrmion formation. **c**, Temperature ($T$) dependence of longitudinal resistivity ($\rho_{xx}$, top panel) and out-of-plane magnetization measured at 0.1 T ($M$, bottom panel) for SRO5/Sub, SIO2/SRO5/Sub, and SRO5/SIO2/Sub. **d**, Anomalous Hall conductance ($\sigma_{AHE}$) as a function of magnetization ($M$). **e-g**, The sum of anomalous and topological Hall resistivity ($\rho_{AHE} + \rho_{THE}$) at 2 K as a function of external magnetic field ($B$) under application of gate bias $V_G = -180$ V (blue lines) and 200 V (red lines). On top, respective sample structure is shown. $\rho_{AHE} + \rho_{THE}$ is deduced by subtracting a $B$-linear ordinary Hall component ($R_H B$) from the Hall resistivity ($\rho_{yx}$). Black arrows indicate the sweep direction of $B$.

**Figure 2 | Electric-field control of anomalous Hall effect**

**a**, Magnetic-field ($B$) dependence of $\rho_{AHE} + \rho_{THE}$ (solid lines) at 30 K under $V_G = -180$ V (left panel), 0 V (middle panel), and 200 V (right panel) for SRO5/SIO2/Sub. Magneto-optic Kerr rotation as a function of $B$ under the same gate bias is also shown by broken lines in each panel. Yellow coloured regions correspond to $\rho_{THE}$. Black arrows indicate the sweep direction of $B$. **b**, Temperature ($T$) dependence of $\rho_{AHE}$ under $V_G = -180$ V (blue), 0 V (grey) and 200 V (red). The black box corresponds to the data in **a**.

**Figure 3 | Electric-field control of topological Hall effect**

**a**, Topological Hall resistivity ($\rho_{THE}$) at 30 K as a function of external magnetic field ($B$) under $V_G = -180$ V (blue line), 0 V (grey line) and 200 V (red line) for SRO5/SIO2/Sub. Black arrows indicate the sweep direction of $B$. **b**, Temperature ($T$) dependence of $\rho_{THE}$ under gate bias. **c**, Colour map of $\rho_{THE}$ in the $T$-$B$ plane under $V_G = -180$ V (left), 0 V (middle) and 200 V (right).

**Figure 4 | Qualitative analyses of modulations in AHE and THE**

**a**, Difference in anomalous Hall resistivity ($\Delta\rho_{AHE}$) between $V_G = 200$ V and $V_G = -180$ V as a function of longitudinal resistivity variation ratio [$\Delta\rho_{xx}/\rho_{xx}$ (0 V)] at temperatures ranging from 2 K to 80 K. **b**, Modulation ratio of topological Hall resistivity [$\Delta\rho_{THE}/\rho_{THE}$ (0 V)] as a function of $\Delta\rho_{xx}/\rho_{xx}$ (0 V). Red dash-dotted line is the calculated variation where gate bias only changes carrier density. Pink dotted line is the guide to the eyes. Broken arrows indicate temperature variation.

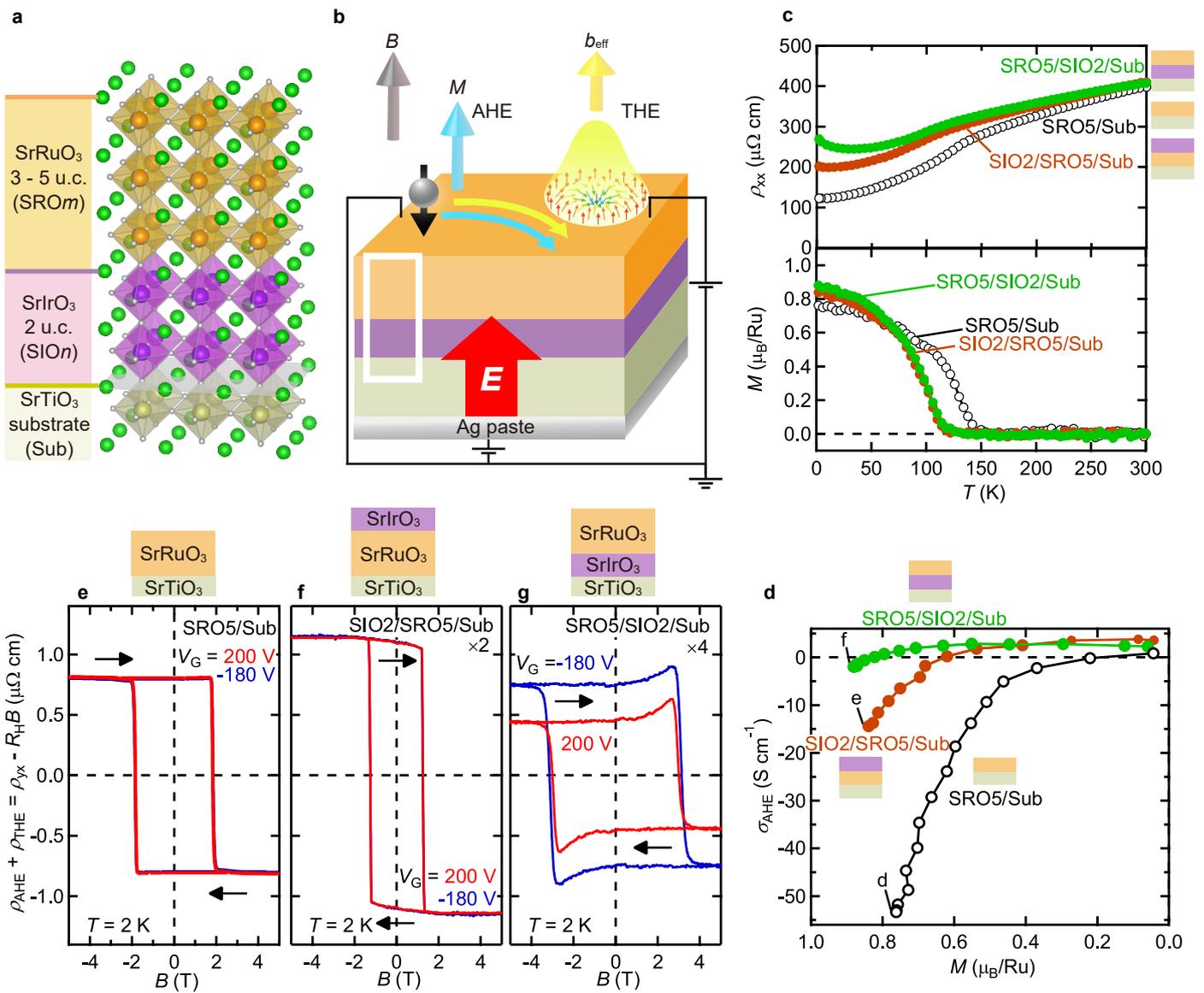

Figure 1 Y. Ohuchi *et al.*

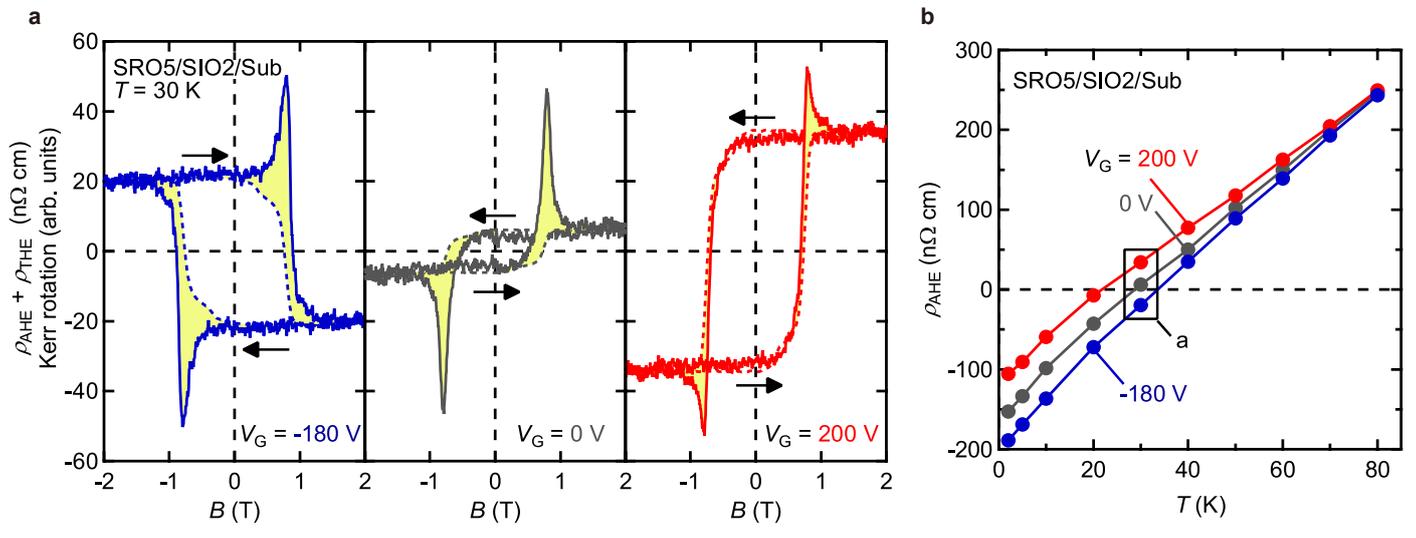

Figure 2 Y. Ohuchi *et al*.

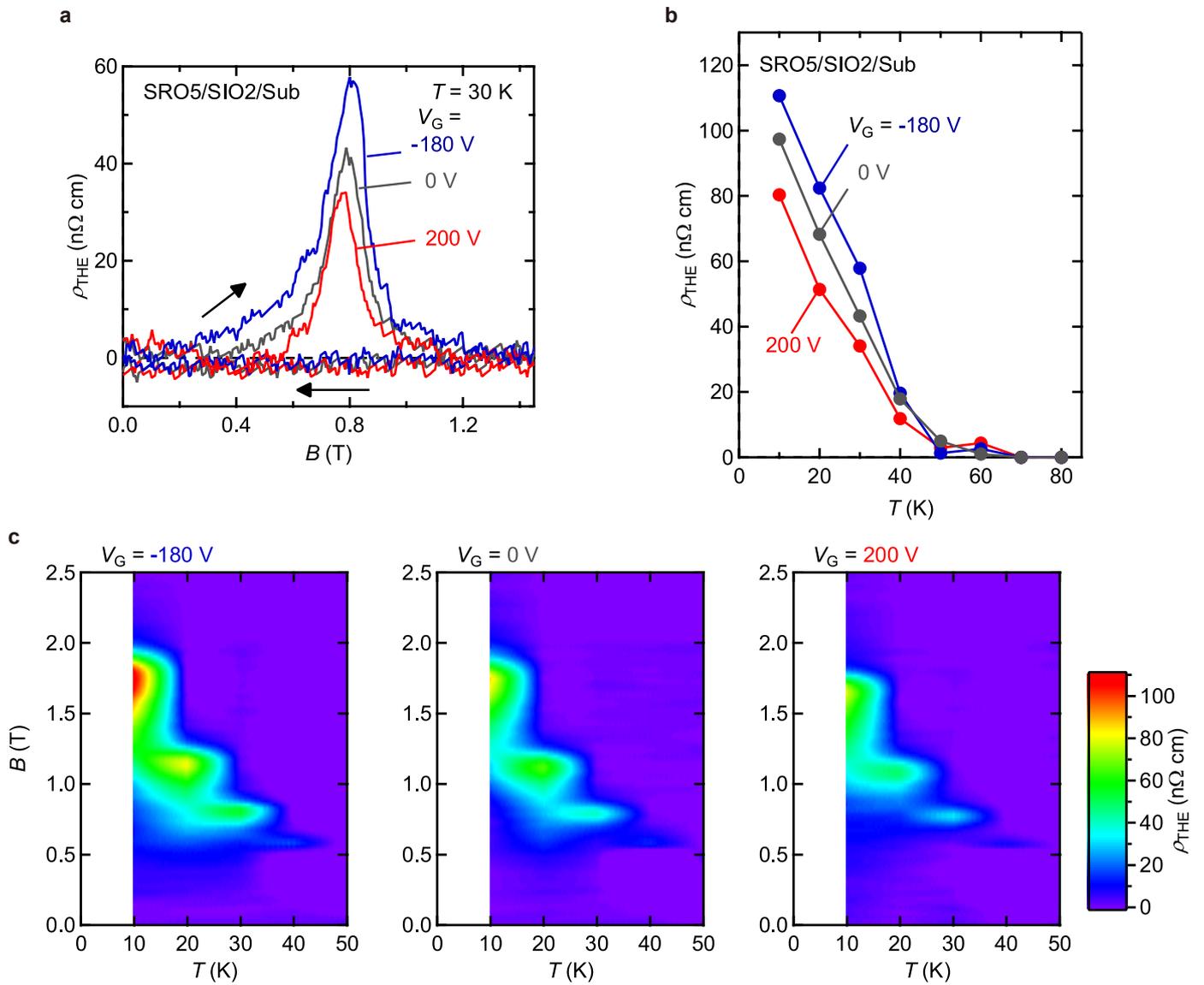

Figure 3 Y. Ohuchi *et al*.

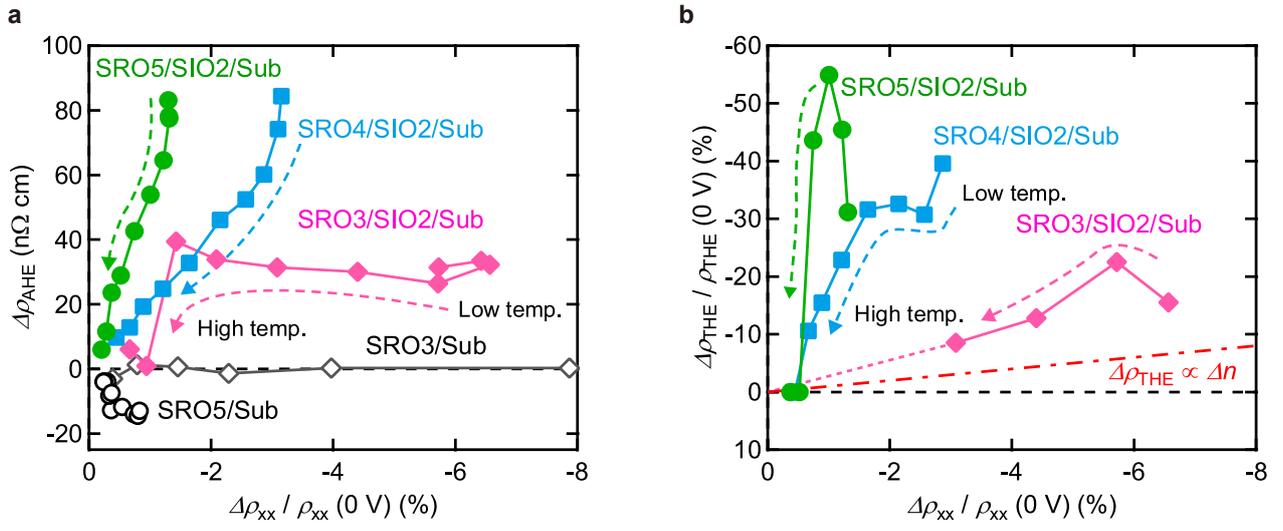

Figure 4 Y. Ohuchi *et al*.